\documentclass{article}
\usepackage{spconf,amsmath,graphicx}
\usepackage{subfig}
\usepackage[colorlinks=true]{hyperref}

\usepackage[table,xcdraw]{xcolor}

\usepackage{multirow}

\usepackage{pifont}

\newcommand\blfootnote[1]{%
  \begingroup
  \renewcommand\thefootnote{}\footnote{#1}%
  \addtocounter{footnote}{-1}%
  \endgroup
}

\title{ErrorNet: A Unified Error Injection-Prediction Framework Applied to Retinal Vessel Segmentation}

\title{ErrorNet: Automated Correction of Segmentation Errors Caused by Unrepresentative Datasets}

\title{ErrorNet: Learning error representations from limited data to improve vascular segmentation}
\name{Nima Tajbakhsh$^{*\dagger}$, Brian Lai$^{*\dagger}$, Shilpa P. Ananth$^{\dagger}$, Xiaowei Ding$^{\dagger}$$^{\ddagger}$}
\address{$^{\dagger}$VoxelCloud, Inc.\\ $^{\ddagger}$ Shanghai Jiao Tong University} %
\begin{document}
\maketitle
\begin{abstract}
Deep convolutional neural networks have proved effective in segmenting lesions and anatomies in various medical imaging modalities. However, in the presence of small sample size and domain shift problems, these models often produce masks with non-intuitive segmentation mistakes. In this paper, we propose a segmentation framework called ErrorNet, which learns to correct these segmentation mistakes through the repeated process of injecting systematic segmentation errors to the segmentation result based on a learned shape prior, followed by attempting to predict the injected error.  During inference, ErrorNet corrects the segmentation mistakes by adding the predicted error map to the initial segmentation result.  ErrorNet has advantages over alternatives based on domain adaptation or CRF-based post processing, because it requires neither domain-specific parameter tuning nor any data from the target domains. We have evaluated ErrorNet using five public datasets for the task of retinal vessel segmentation. The selected datasets differ in size and patient population, allowing us to evaluate the effectiveness of ErrorNet in handling small sample size and domain shift problems. Our experiments demonstrate that ErrorNet outperforms a base segmentation model, a CRF-based post processing scheme, and a domain adaptation method, with a greater performance gain in the presence of the aforementioned dataset limitations. 

\end{abstract}
\begin{keywords}
retinal vessel segmentation, limited data, domain shift, error prediction, error correction
\end{keywords}
%

\blfootnote{$^*$ Authors contributed equally}
\section{Introduction}

Medical imaging datasets are often unrepresentative of the patient population, lacking adequate annotated images or otherwise being limited to a particular clinical site or vendor. The former leads to the small sample size problem whereas the latter causes the domain shift problem. In the presence of unrepresentative training datasets, even the most sophisticated architectures (e.g., \cite{zhou2018unet++,imran2019automatic}) may generate non-intuitive segmentation errors such as holes in the segmentations of connected organs or breaks along the segmented vessels. Although acquiring additional annotations strikes as a natural workaround to reduce systematic segmentation errors caused by unrepresentative datasets, it incurs substantial annotation cost. Recent active learning and interactive segmentation methodologies \cite{sourati2019intelligent,kuo2018cost,sakinis2019interactive} provide cost-effective solutions to expand medical datasets, but they still require highly-trained medical experts in the loop. Unsupervised domain adaptation is an expert-free solution that aims to expand medical datasets by bridging the domain shift between the current training set and the target test sets. However, this approach requires unlabeled data from the target domains, which not only is scarcely available, but also does not scale well to many target domains (e.g., widespread clinical deployment). Post-processing methods based on conditional random fields (CRFs) is another approach to reducing systematic segmentation errors caused by limited datasets. While effective in natural images,  application of CRFs in medical images have shown inconclusive results \cite{tajbakhsh2019embracing}. Furthermore, post-processing with CRFs often requires extensive application-specific parameter tuning. There is a need to develop a methodology that can reduce the systematic errors caused by dataset limitations without requiring experts, additional datasets, or extensive parameter tuning.

\begin{figure*}[!ht]
\centering
    \includegraphics[width=0.99\linewidth]{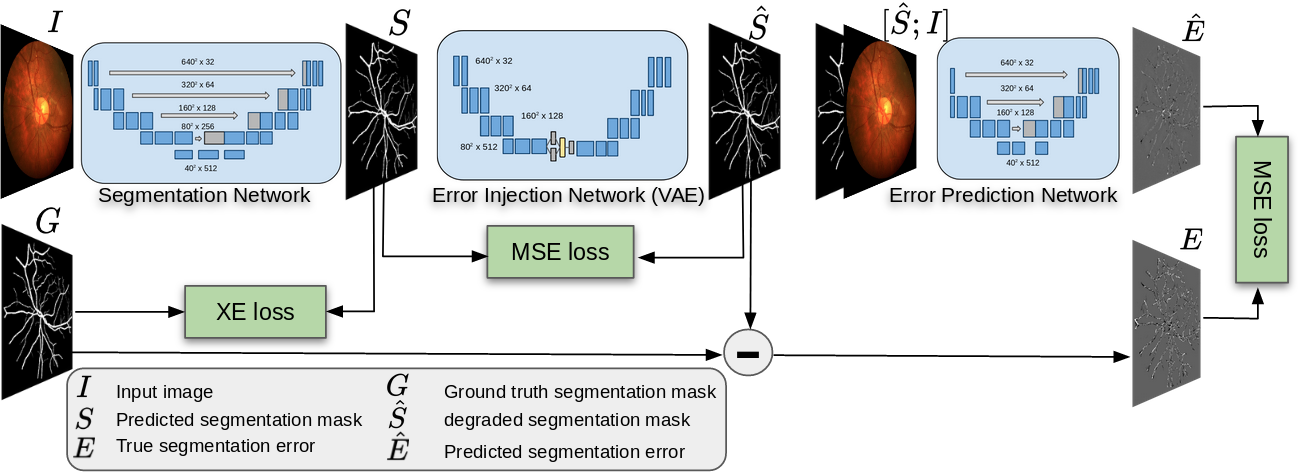}
    \caption{Overview of the suggested segmentation framework, ErrorNet. Given a training image $I$, the segmentation network generates the initial segmentation result $S$, which is then degraded by the error injection network based on a learned shape prior of the vessels, resulting in the degraded segmentation result $\hat{S}$. The prediction network takes as input the original image stacked with the degraded segmentation result, and outputs an error map $\hat{E}$, which attempts to predict the true error, $E$, between the ground truth and the degraded segmentation map at the pixel-level.  \iffalse To avoid clutter, KL-divergence loss of VAE is not shown.\fi}
    \label{fig:method}
\end{figure*}

In this paper, we propose ErrorNet, a segmentation framework with the capability of reducing segmentation errors caused by domain shifts or limited datasets. 
ErrorNet consists of a base segmentation network, an error injection network, and an error prediction network. During training, the error injection network degrades the segmentation result from the base network according to the shape prior of the region of interest. The degraded segmentation is then fed to the error prediction network that aims to estimate an error map between the degraded input and ground truth. Essentially, the segmentation network and error injection network work together to prepare a diverse set of input data for training the error prediction network. At test time, we feed the segmentation results from the base network to the error prediction network directly, and then obtain the corrected segmentation result by adding the predicted error map to the initial segmentation result. We have evaluated ErrorNet for the task of retinal vessel segmentation in fundus images using five public datasets. This choice of datasets allows us to assess the error correction capability of ErrorNet in both same-domain and cross-domain evaluation settings. Our experiments demonstrate that ErrorNet outperforms a base segmentation model, a CRF-based post processing scheme, and a domain adaptation method, with a greater performance gain for smaller training datasets that face larger domain shifts.

\noindent {\bf \underline{Contributions.}} Our contributions include: 1) a novel segmentation framework with an error correction mechanism based on shape prior, 2) extensive evaluation using five public datasets in both same- and cross-dataset evaluation, 3) demonstrated gain over several  baselines including CRF-based post processing and a domain adaptation method.

\section{Related work}
Due to limited space, we contrast our approach against domain adaptation, post-segmentation refinement, and conditional prior networks, and refer the readers to \cite{tajbakhsh2019embracing} for a comprehensive survey on methods for medical image segmentation with limited datasets.

\vspace{4pt}
\noindent\underline{\emph{Domain adaptation:}} These methods typically require unlabeled \cite{huo2018synseg,chen2019synergistic} or often labeled data \cite{dou2018pnp,zhang2018translating} from the target domains. Thus, they are hardly applicable when the target domain is unknown a priori or too diverse to have data from. In contrast, ErrorNet does not require data from the target domains as it corrects segmentation errors based on a shape prior learned during the training stage.

\vspace{4pt}
\noindent\underline{\emph{Conditional Prior Networks (CPNs)}:}
CPNs can faithfully learn a shape prior for flows between images \cite{yang2019dense}. We adopt a similar structure but instead of learning a prior distribution of flows, we use the CPN architecture to learn the prior distribution of segmentation masks. We additionally extend CPNs such that they can generate examples that lie within the learned segmentation prior \cite{kingma2013auto}. We learn the shape prior of segmentation masks so that ErrorNet can leverage the learned segmentation shape prior and train itself to correct imperfect segmentation masks when only limited data is available. 

\vspace{4pt}
\noindent\underline{\emph{Post-processing schemes:}} Methods based on different variants of CRFs can be used to force connectivity constraints in segmentation results \cite{kamnitsas2017efficient,wachinger2018deepnat}. However, these methods often require extensive parameter tuning and have shown only mixed results for medical image segmentation \cite{tajbakhsh2019embracing}. In contrast, ErrorNet is end-to-end trainable and application agnostic, eliminating the need for heuristic designs. Denoising autoencoders have also been used as post-processing recently \cite{larrazabal2019anatomical}, but handcrafted, domain-specific error patterns are required for training. This limitation is overcome in ErrorNet as the error patterns are learned systematically.

\section{Method}

\figurename~\ref{fig:method} shows the overview of our proposed segmentation framework, ErrorNet, which consists of three consecutive networks: a base segmentation network, an error injection network, and an error prediction network. We explain each individual network as follows:

\vspace{4pt}
\noindent\underline{\emph{Base segmentation network:}} We choose the widely used U-Net as the base segmentation network, which is trained by minimizing the cross entropy loss, $L_{seg}=-\sum_{i}log(p_i)$.

\vspace{4pt}
\noindent\underline{\emph{Error injection network:}}
The task for the error injection network is to degrade the segmentation result  by injecting error patterns to the initial segmentation result. However, it is critical for the error patterns to be representative; otherwise, the subsequent error prediction network would learn an unrelated and perhaps trivial vision task, leading to an ineffective error correction mechanism. Also, the error patterns must be diverse; otherwise, the  subsequent error prediction network will overfit to a limited set of error patterns in the training set, particularly when the training set is small. The importance of diverse error patterns is even more pronounced for cross-domain model evaluation where the base segmentation model may produce segmentation maps with error patterns that are partially or largely different from that of the source training dataset. The choice of error injection network is thus critical to the success of the suggested framework. 

For this purpose, we use a variational autoencoder (VAE), which is trained by minimizing $ L_{vae} = \sum_{i} (\hat{S}^{(i)} - S^{(i)})^2+KL(P_{\theta^*}(z^{(i)}|S^{(i)})||N(0,1))$ where the first term constrains the degraded mask to be similar to the initial segmentation result and the second term constraints the latent space of the VAE to follow a standard normal distribution. A VAE generates representative and diverse error patterns, because, during training it learns a distribution over segmentation maps in a high dimensional feature space. By sampling from this distribution, VAE can generate diverse variants of a given segmentation map, thereby increasing the input space of imperfect segmentations, enabling us to train a more robust error prediction network. 
 
\vspace{4pt}
\noindent\underline{\emph{Error prediction network:}}
We use a shallow U-Net for the task of error prediction, which takes as input the degraded segmentation map stacked with the original image, and returns a k-channel error prediction map at the original image resolution, $\hat{E}^{(i)}$, where $k$ is the number of classes. We use the hyperbolic tangent
function in the output layer, because its output changes between -1 and 1, allowing the class corrections in both positive (strengthening) and negative (weakening) directions. We train this network by minimizing $L_{pred} = \sum_{i}(\hat{E}^{(i)} - E^{(i)})$ where $E^{(i)}=(\hat{S}^{(i)}-G^{(i)})^2$ with $G^{(i)}$ being the ground truth mask for the $i^{th}$ image in the batch. 

\vspace{4pt}
During testing, we bypass the error injection network, directly passing the initial segmentation result to the error prediction model. The final segmentation mask is obtained by adding the predicted error map to the initial segmentation.

\section{Experiments}

\noindent\underline{\emph{Architecture details:}}
We use a U-Net as the base segmentation model. The U-Net consists of 4 downsampling blocks in the encoder and 4 upsampling blocks in the decoder. All convolutions layers use 3x3 kernels. We have followed the best practices as suggested in \cite{isensee18} for configuring and training our U-Net architecture. Specifically, we have used instance normalization in downsampling blocks and leaky Relu as activation functions, and further excluded large black background regions while normalizing the training images, to name a few. For the error injection module, we use a VAE with 3 downsampling blocks, a 6400 dimensional latent feature space, and 3 upsampling blocks. To ensure that the injected error patterns do not transform the segmentation mask completely, we sample from the latent space with a variance of 0.0001. The error prediction module follows a U-Net architecture with 3 downsampling blocks followed by 3 upsampling blocks with skip connections. Both VAE and error prediction network use batch normalization and relu activation functions. The number of kernels in the VAE and error prediction network were optimized so that a GPU with 12 GB RAM can hold the entire ErrorNet in memory. We refer the readers to the \href{https://arxiv.org/abs/1910.04814}{appendix} for architecture details.

\vspace{4pt}
\noindent\underline{\emph{Training Details:}}
While ErrorNet can be trained end-to-end, we have found that stage-wise training facilitates convergence. Specifically, we first train the base segmentation by minimizing the segmentation loss function, $L = L_{seg}$. We then freeze the weights of the base segmentation network and train error injection network by minimizing the VAE loss, $L = L_{vae}$. Once VAE is trained, we train the error prediction network by minimizing $L = L_{pred}$, while freezing the weights of the base segmentation and the error injection networks. We refer to this training scheme as \emph{stage-wise} training henceforth. The ErrorNet can now be trained jointly in an end-to-end fashion, while freezing the weights of the error injection module, effectively minimizing $L = L_{pred} + L_{seg}$. We refer to this training scheme as \emph{Joint Tr} in Table~\ref{tab:results}.

\vspace{4pt}
\noindent\underline{\emph{Datasets:}}
Table~\ref{tab:dataset} summarizes the 5 datasets used to evaluate  ErrorNet for the task of retinal vessel segmentation. The selected datasets vary in terms of size, population, and acquisition machine, allowing us to evaluate the effectiveness of ErrorNet under different sample sizes and domain shift.

\begin{table}[]
\caption{Datasets used in our experiments}
\label{tab:dataset}
\resizebox{0.95\columnwidth}{!}{%
\begin{tabular}{lllllllll}
\hline
\multicolumn{1}{c}{Dataset} &  & \multicolumn{3}{c}{\# images} &  & \multicolumn{3}{c}{Data splits} \\ \hline
\multicolumn{1}{c}{} &  & \multicolumn{1}{c}{Total} & \multicolumn{1}{c}{Diseased} & \multicolumn{1}{c}{Healthy} &  & \multicolumn{1}{c}{Train} & \multicolumn{1}{c}{Val} & \multicolumn{1}{c}{Test} \\ \hline \hline
DRIVE &  & 40 & 7 & 33 &  & 18 & 2 & 20 \\ \hline
STARE &  & 20 & 10 & 10 &  & 10 & 2 & 8 \\ \hline
CHASE &  & 20 & 0 & 20 &  & 17 & 5 & 6 \\ \hline
ARIA &  & 143 & 82 & 61 &  & 121 & 5 & 17 \\ \hline
HRF &  & 45 & 30 & 15 &  & 26 & 5 & 14 \\ \hline
\end{tabular}%
}
\end{table}

\begin{table*}[t]
\caption{\small Comparison between ErrorNet and other performance baselines. Dice is used for comparison (see the \href{https://arxiv.org/abs/1910.04814}{appendix} for IoU). Grey columns indicate same-domain evaluation whereas the other columns contain the results for cross-domain evaluation. ErrorNet outperforms the baselines on-average, with a larger gain in the presence of domain shift (cross-domain evaluation) and small sample size (the small \emph{Chase} dataset used for training). Ablation studies show that VAE and joint-training are effective in improving the performance of ErrorNet.}
\label{tab:results}
\resizebox{\textwidth}{!}{%
\begin{tabular}{|l|l|l|l|lllllll|lllllll|}
\hline
\multicolumn{1}{|c|}{} & \rotatebox[origin=c]{90}{Err Pred} & \rotatebox[origin=c]{90}{VAE} & \rotatebox[origin=c]{90}{Joint Tr} & \multicolumn{1}{c}{Train on $\rightarrow$} & \multicolumn{5}{c}{CHASE} &  & \multicolumn{1}{c}{} & \multicolumn{5}{c}{ARIA} &  \\ \cline{2-18} 
\multicolumn{1}{|c|}{\multirow{-2}{*}{Architecture}} &  &  &  & \multicolumn{1}{c}{Test on $\rightarrow$} & \multicolumn{1}{c}{\cellcolor[HTML]{EFEFEF}CHASE} & \multicolumn{1}{c}{DRIVE} & \multicolumn{1}{c}{ARIA} & \multicolumn{1}{c}{STARE} & \multicolumn{1}{c|}{HRF} & Avg. & \multicolumn{1}{c}{} & \multicolumn{1}{c}{CHASE} & \multicolumn{1}{c}{DRIVE} & \multicolumn{1}{c}{\cellcolor[HTML]{EFEFEF}ARIA} & \multicolumn{1}{c}{STARE} & \multicolumn{1}{c|}{HRF} & Avg. \\ \hline
U-Net \cite{isensee18} &  &  &  &  & \cellcolor[HTML]{EFEFEF}79.3 & 67.6 & 60.3 & 59.5 & \multicolumn{1}{l|}{61.5} & 65.6 &  & 76.7 & 77.3 & \cellcolor[HTML]{EFEFEF}72.0 & 71.28 & \multicolumn{1}{l|}{72.3} & 73.9 \\ \hline
U-Net \cite{isensee18} + CRF &  &  &  &  & \cellcolor[HTML]{EFEFEF}81.2 & 65.4 & 62.6 & 56.4 & \multicolumn{1}{l|}{63.6} & 65.8 &  & {\bf 78.4} & 69.5 & \cellcolor[HTML]{EFEFEF}73.0 & 64.6 & \multicolumn{1}{l|}{{\bf 73.5}} & 71.8 \\ \hline
V-GAN\cite{son2017retinal} &  &  &  &  & \cellcolor[HTML]{EFEFEF}79.7 & 71.5 & 64.2 & 61.0 & \multicolumn{1}{l|}{66.4} & 68.5 &  & 68.7 & 75.8 & \cellcolor[HTML]{EFEFEF}69.9 & 66.2 & \multicolumn{1}{l|}{69.3} & 70.0 \\ \hline
DA-ADV \cite{dong2018unsupervised} &  &  &  &  & \cellcolor[HTML]{EFEFEF}72.3 & 69.3 & {\bf 68.2} & 64.7 & \multicolumn{1}{l|}{67.4} & 68.4 &  & 71.5 & 72.9 & \cellcolor[HTML]{EFEFEF}{\bf 73.2} & 71.3 & \multicolumn{1}{l|}{70.7} & 71.9 \\ \hline \hline 
 & \ding{51} &  &  &  & \cellcolor[HTML]{EFEFEF}80.2 & 68.6 & 60.7 & 60.2 & \multicolumn{1}{l|}{62.7} & 66.4 &  & 76.8 & 72.1 & \cellcolor[HTML]{EFEFEF}72.0 & 71.9 & \multicolumn{1}{l|}{72.2} & 73.0 \\ \cline{2-18} 
 & \ding{51} & \ding{51} &  &  & \cellcolor[HTML]{EFEFEF}80.1 & 71.8 & 61.1 & 59.8 & \multicolumn{1}{l|}{67.2} & 68.0 &  & 76.2 & 77.3 & \cellcolor[HTML]{EFEFEF}72.2 & 72.2 & \multicolumn{1}{l|}{72.8} & 74.1 \\ \cline{2-18} 
\multirow{-3}{*}{ErrorNet w/ ablation} & \ding{51} & \ding{51} & \ding{51} &  & \cellcolor[HTML]{EFEFEF}{\bf 81.5} & {\bf 73.2} & 66.5 & {\bf 65.2} & \multicolumn{1}{l|}{{\bf 68.6}} & {\bf 71.0} &  & 76.7 & {\bf 78.9} & \cellcolor[HTML]{EFEFEF}72.0 & {\bf 74.0} & \multicolumn{1}{l|}{72.6} & {\bf 74.8} \\ \hline
\end{tabular}%
}
\end{table*}
\vspace{4pt}
\noindent\underline{\emph{Performance baselines:}} We compare ErrorNet against 1) a U-Net carefully-optimized according to the best practices suggested in \cite{isensee18}, the same U-Net with CRF-based post-processing, a recent unsupervised domain adaptation method \cite{dong2018unsupervised}, and V-GAN \cite{son2017retinal}, which is a modern retinal vessel segmentation network trained in an adversarial manner. 

\vspace{4pt}
\noindent\underline{\emph{Ablation study:}} We compare ErrorNet with and without VAE to study the impact of the error injection network. Without VAE, the error prediction network only sees the error patterns in the training dataset. To study the effect of joint training, we also compare ErrorNet with and without joint training.

\vspace{4pt}
\noindent\underline{\emph{Evaluation scenarios:}} We evaluate ErrorNet in the presence of small sample size and domain shift problems. To study the small size, we train ErrorNet using \emph{Chase}, which is a small dataset, and \emph{Aria}, which is the largest dataset under study. To study the domain shift problem, we evaluate the models above on the datasets other than the one they are trained on.

\vspace{4pt}
\noindent\underline{\emph{Results:}} Table~\ref{tab:results} summarizes the results of both evaluation scenarios. When \emph{Chase} dataset is used for training and testing, ErrorNet achieves a Dice of 81.5 outperforming all performance baselines. The ablation study also shows that ErrorNet with joint training achieves a 1-point increase in Dice. Inclusion of VAE, on the other hand, shows no significant performance gain. This is because the training and test domains are the same (\emph{Chase}). In the cross-domain evaluation, ErrorNet and in particular the VAE module achieve greater performance gains over the baselines.  Specifically, ErrorNet achieves an average Dice score of 71.0, outperforming the second best \cite{son2017retinal} and third best method \cite{dong2018unsupervised} by 2.4 and 2.5 points, respectively. The VAE module also enables a 1.6-point increase in Dice. The widened performance gap is due to the domain shift caused by different patient population and pathologies present in the datasets. While \emph{Chase} contains only healthy fundus images of children's eyes with a central vessel reflex, all the other datasets used for testing contain pathological cases from adult populations. We hypothesize that  ErrorNet  can effectively learn the general structure of eye vessels; and thus, it can help correct mis-segmentations introduced by dataset limitations. 

ErrorNet trained with \emph{Aria} continues to improve the segmentation performance over the baseline models, with both VAE and joint training features showing consistent performance gains. However, the superiority over baselines is not as drastic as in the case where ErrorNet was trained with \emph{Chase}. Intuitively, these results make sense. \emph{Aria} is a larger, more varied dataset with images from both healthy and diseased patients; and thus, the models trained on \emph{Aria} generalize better to other datasets. As a result, the improvements made by the error correction module are smaller.

\begin{figure}[t]
\centering
\includegraphics[width=0.99\linewidth, keepaspectratio]{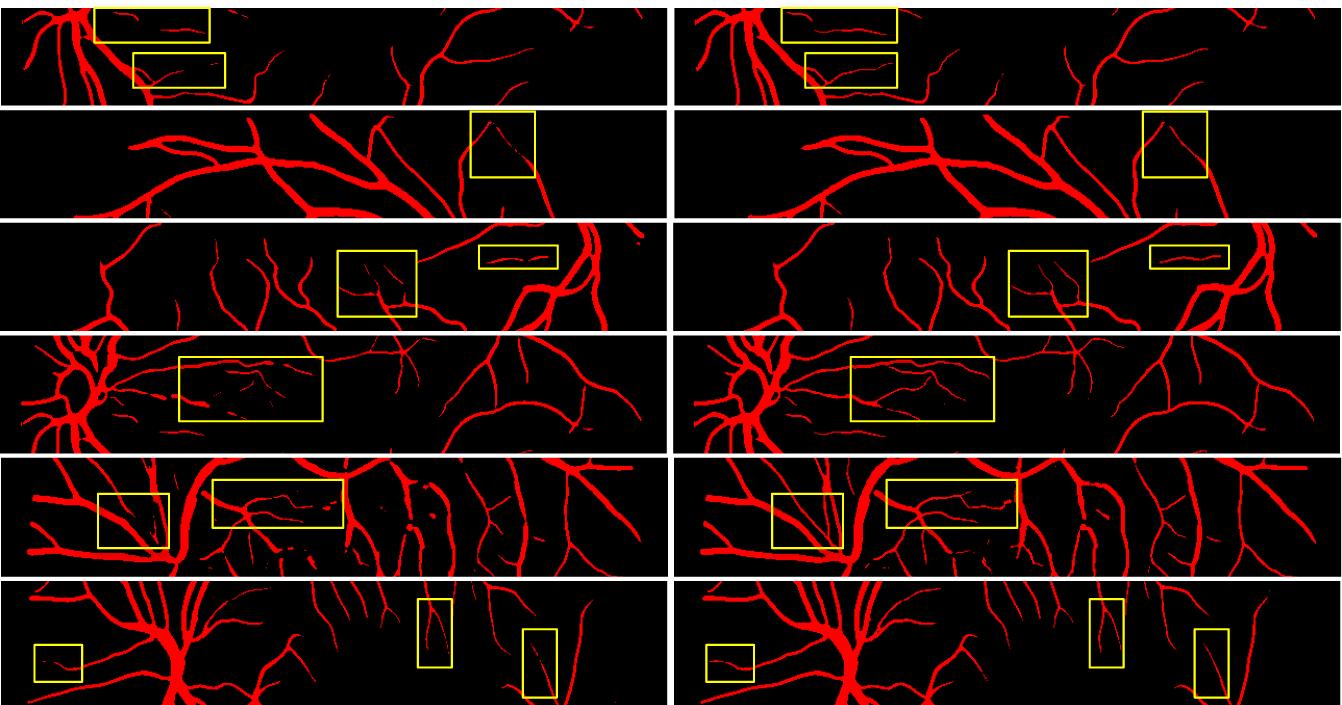}
    \caption{\small 
    ErrorNet is effective in bridging breaks along segmented vessels. Each row compares the segmentation results before (left) and after (right) error correction. The yellow boxes indicate regions where the ErrorNet has connected fragmented vessels or sharpened vessel structures.Full-image results are available in the \href{https://arxiv.org/abs/1910.04814}{appendix}.
    } 
    \label{fig:qual_comp}
\end{figure}

\vspace{4pt}
\noindent\underline{\emph{Qualitative comparison:}}
\figurename~\ref{fig:qual_comp} compares the segmentation results before and after error correction by the error prediction network. Recall that the error injection network is not used during inference---segmentation results are directly sent to the error prediction network for error correction. As highlighted by the yellow boxes, ErrorNet has connected fragmented vessels or sharpened vessel structures. 

\section{Conclusion}
We presented ErrorNet, a framework for systematic handling of segmentation errors caused by limited datasets. We evaluated ErrorNet using 5 public datasets for the task of retinal vessel segmentation. Our results demonstrated the effectiveness of ErrorNet in both same-dataset and cross-dataset evaluations, particularly when the size of training set was small and domain shift was large. Our future work would focus on evaluating ErrorNet on other medical image segmentation tasks as well as evaluating the effectiveness of ErrorNet for the task of active learning.

\bibliographystyle{IEEEbib}
\bibliography{refs}

\begin{thebibliography}{10}

\bibitem{zhou2018unet++}
Zongwei Zhou, Md~Mahfuzur~Rahman Siddiquee, Nima Tajbakhsh, and Jianming Liang,
\newblock ``Unet++: A nested u-net architecture for medical image
  segmentation,''
\newblock in {\em Deep Learning in Medical Image Analysis and Multimodal
  Learning for Clinical Decision Support}, pp. 3--11. Springer, 2018.

\bibitem{imran2019automatic}
Abdullah-Al-Zubaer Imran, Ali Hatamizadeh, Shilpa~P Ananth, Xiaowei Ding,
  Demetri Terzopoulos, and Nima Tajbakhsh,
\newblock ``Automatic segmentation of pulmonary lobes using a progressive dense
  v-network,''
\newblock {\em arXiv preprint arXiv:1902.06362}, 2019.

\bibitem{sourati2019intelligent}
Jamshid Sourati, Ali Gholipour, Jennifer~G Dy, Xavier Tomas-Fernandez, Sila
  Kurugol, and Simon~K Warfield,
\newblock ``Intelligent labeling based on fisher information for medical image
  segmentation using deep learning,''
\newblock {\em IEEE transactions on medical imaging}, 2019.

\bibitem{kuo2018cost}
Weicheng Kuo, Christian H{\"a}ne, Esther Yuh, Pratik Mukherjee, and Jitendra
  Malik,
\newblock ``Cost-sensitive active learning for intracranial hemorrhage
  detection,''
\newblock in {\em International Conference on Medical Image Computing and
  Computer-Assisted Intervention}. Springer, 2018, pp. 715--723.

\bibitem{sakinis2019interactive}
Tomas Sakinis, Fausto Milletari, Holger Roth, Panagiotis Korfiatis, Petro
  Kostandy, Kenneth Philbrick, Zeynettin Akkus, Ziyue Xu, Daguang Xu, and
  Bradley~J Erickson,
\newblock ``Interactive segmentation of medical images through fully
  convolutional neural networks,''
\newblock {\em arXiv preprint arXiv:1903.08205}, 2019.

\bibitem{tajbakhsh2019embracing}
Nima Tajbakhsh, Laura Jeyaseelan, Qian Li, Jeffrey Chiang, Zhihao Wu, and
  Xiaowei Ding,
\newblock ``Embracing imperfect datasets: A review of deep learning solutions
  for medical image segmentation,''
\newblock {\em arXiv preprint arXiv:1908.10454}, 2019.

\bibitem{huo2018synseg}
Yuankai Huo, Zhoubing Xu, Hyeonsoo Moon, Shunxing Bao, Albert Assad, Tamara~K
  Moyo, Michael~R Savona, Richard~G Abramson, and Bennett~A Landman,
\newblock ``Synseg-net: Synthetic segmentation without target modality ground
  truth,''
\newblock {\em IEEE transactions on medical imaging}, vol. 38, no. 4, pp.
  1016--1025, 2018.

\bibitem{chen2019synergistic}
Cheng Chen, Qi~Dou, Hao Chen, Jing Qin, and Pheng-Ann Heng,
\newblock ``Synergistic image and feature adaptation: Towards cross-modality
  domain adaptation for medical image segmentation,''
\newblock {\em arXiv preprint arXiv:1901.08211}, 2019.

\bibitem{dou2018pnp}
Qi~Dou, Cheng Ouyang, Cheng Chen, Hao Chen, Ben Glocker, Xiahai Zhuang, and
  Pheng-Ann Heng,
\newblock ``Pnp-adanet: Plug-and-play adversarial domain adaptation network
  with a benchmark at cross-modality cardiac segmentation,''
\newblock {\em arXiv preprint arXiv:1812.07907}, 2018.

\bibitem{zhang2018translating}
Zizhao Zhang, Lin Yang, and Yefeng Zheng,
\newblock ``Translating and segmenting multimodal medical volumes with
  cycle-and shape-consistency generative adversarial network,''
\newblock in {\em Proceedings of the IEEE Conference on Computer Vision and
  Pattern Recognition}, 2018, pp. 9242--9251.

\bibitem{yang2019dense}
Yanchao Yang, Alex Wong, and Stefano Soatto,
\newblock ``Dense depth posterior (ddp) from single image and sparse range,''
\newblock in {\em Proceedings of the IEEE Conference on Computer Vision and
  Pattern Recognition}, 2019, pp. 3353--3362.

\bibitem{kingma2013auto}
Diederik~P Kingma and Max Welling,
\newblock ``Auto-encoding variational bayes,''
\newblock {\em arXiv preprint arXiv:1312.6114}, 2013.

\bibitem{kamnitsas2017efficient}
Konstantinos Kamnitsas, Christian Ledig, Virginia~FJ Newcombe, Joanna~P
  Simpson, Andrew~D Kane, David~K Menon, Daniel Rueckert, and Ben Glocker,
\newblock ``Efficient multi-scale 3d cnn with fully connected crf for accurate
  brain lesion segmentation,''
\newblock {\em Medical image analysis}, vol. 36, pp. 61--78, 2017.

\bibitem{wachinger2018deepnat}
Christian Wachinger, Martin Reuter, and Tassilo Klein,
\newblock ``Deepnat: Deep convolutional neural network for segmenting
  neuroanatomy,''
\newblock {\em NeuroImage}, vol. 170, pp. 434--445, 2018.

\bibitem{larrazabal2019anatomical}
Agostina~J. Larrazabal, Cesar Martinez, and Enzo Ferrante,
\newblock ``Anatomical priors for image segmentation via post-processing with
  denoising autoencoders,''
\newblock in {\em Medical Image Computing and Computer Assisted Intervention --
  MICCAI 2019}. 2019, pp. 585--593, Springer International Publishing.

\bibitem{isensee18}
Fabian Isensee, Philipp Kickingereder, Wolfgang Wick, Martin Bendszus, and
  Klaus~H Maier-Hein,
\newblock ``No new-net,''
\newblock in {\em International MICCAI Brainlesion Workshop}. Springer, 2018,
  pp. 234--244.

\bibitem{son2017retinal}
Jaemin Son, Sang~Jun Park, and Kyu-Hwan Jung,
\newblock ``Retinal vessel segmentation in fundoscopic images with generative
  adversarial networks,''
\newblock {\em arXiv preprint arXiv:1706.09318}, 2017.

\bibitem{dong2018unsupervised}
Nanqing Dong, Michael Kampffmeyer, Xiaodan Liang, Zeya Wang, Wei Dai, and Eric
  Xing,
\newblock ``Unsupervised domain adaptation for automatic estimation of
  cardiothoracic ratio,''
\newblock in {\em International Conference on Medical Image Computing and
  Computer-Assisted Intervention}. Springer, 2018, pp. 544--552.

\end{thebibliography}

\setcounter{figure}{0}\renewcommand{\thefigure}{A.\arabic{figure}} 
\setcounter{table}{0}\renewcommand{\thetable}{A.\arabic{table}} 
\newpage
\section*{Appendix}
This appendix consists of 6 figures and 4 tables. The figures serve to illustrate how the error correction mechanism of ErrorNet improves the segmentation results in high resolution uncropped images. Note that, due to limited space, we showed only low resolution cropped results in the main text. The tables contain our segmentation results based on IoU and also architecture details for the base segmentation network, error injection network, and error prediction network. The readers are welcome to contact Nima Tajbakhsh at ntajbakhsh@voxelcloud.io for further clarification on our method, results, or architecture details.


\begin{figure*}[h]
\centering
    \includegraphics[width=0.99\linewidth, keepaspectratio]{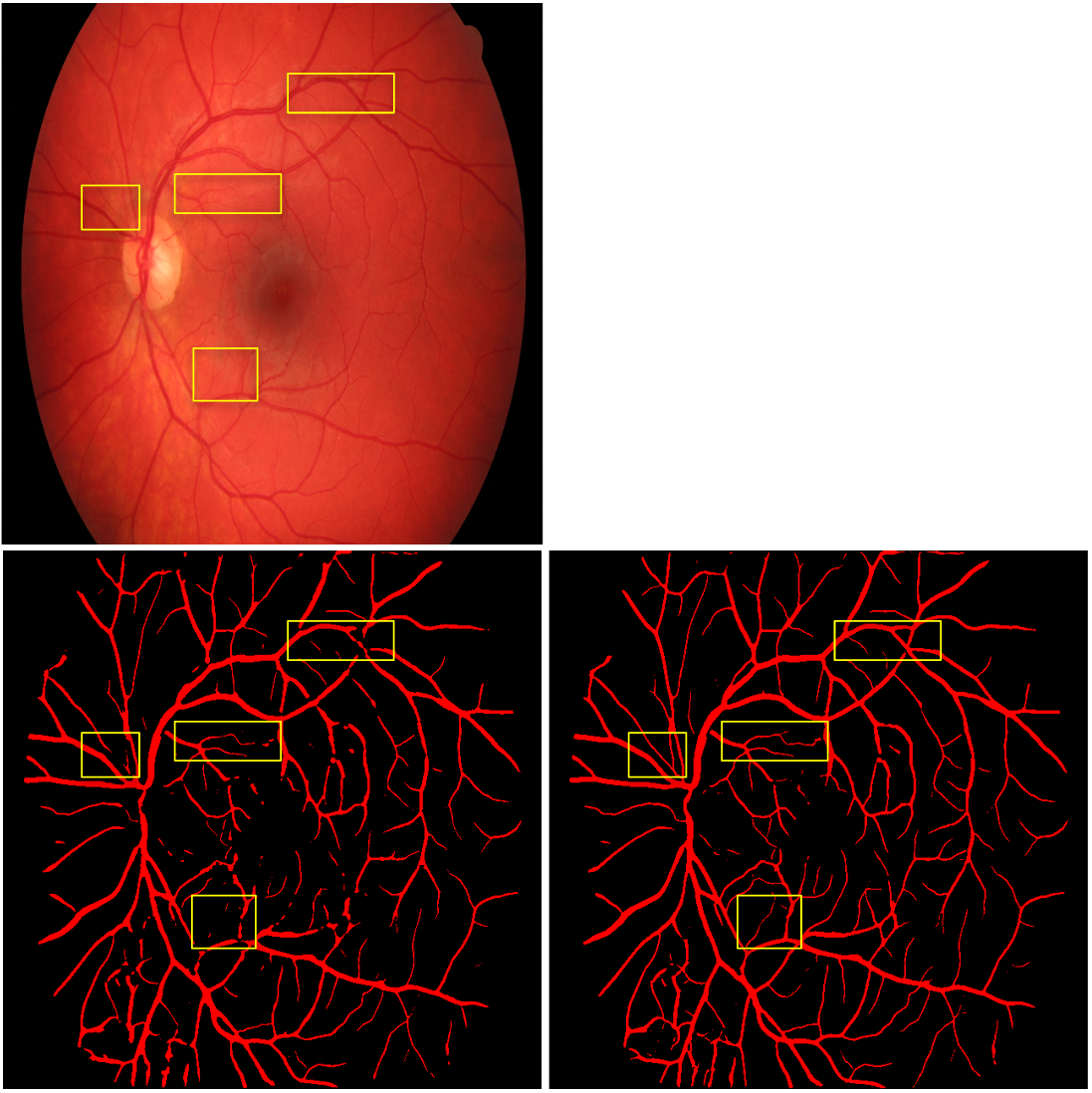}
    \caption{{\color{blue}[\emph{Chase} $\rightarrow$ \emph{HRF}]} Effectiveness of ErrorNet for cross-dataset evaluation, where the training set comes from the \emph{Chase} dataset but the test set comes from the \emph{HRF} dataset. Top: Fundus image. Bottom-Left: Segmentation result for an \emph{HRF} dataset image from the segmentation network (before error correction). Bottom-Right: corresponding segmentation result generated by ErrorNet after error correction. The yellow boxes indicate example regions where the ErrorNet model has connected fragmented vessels or sharpened vessel structures. } 
\end{figure*}

\begin{figure*}[h]
\centering
    \includegraphics[width=0.99\linewidth, keepaspectratio]{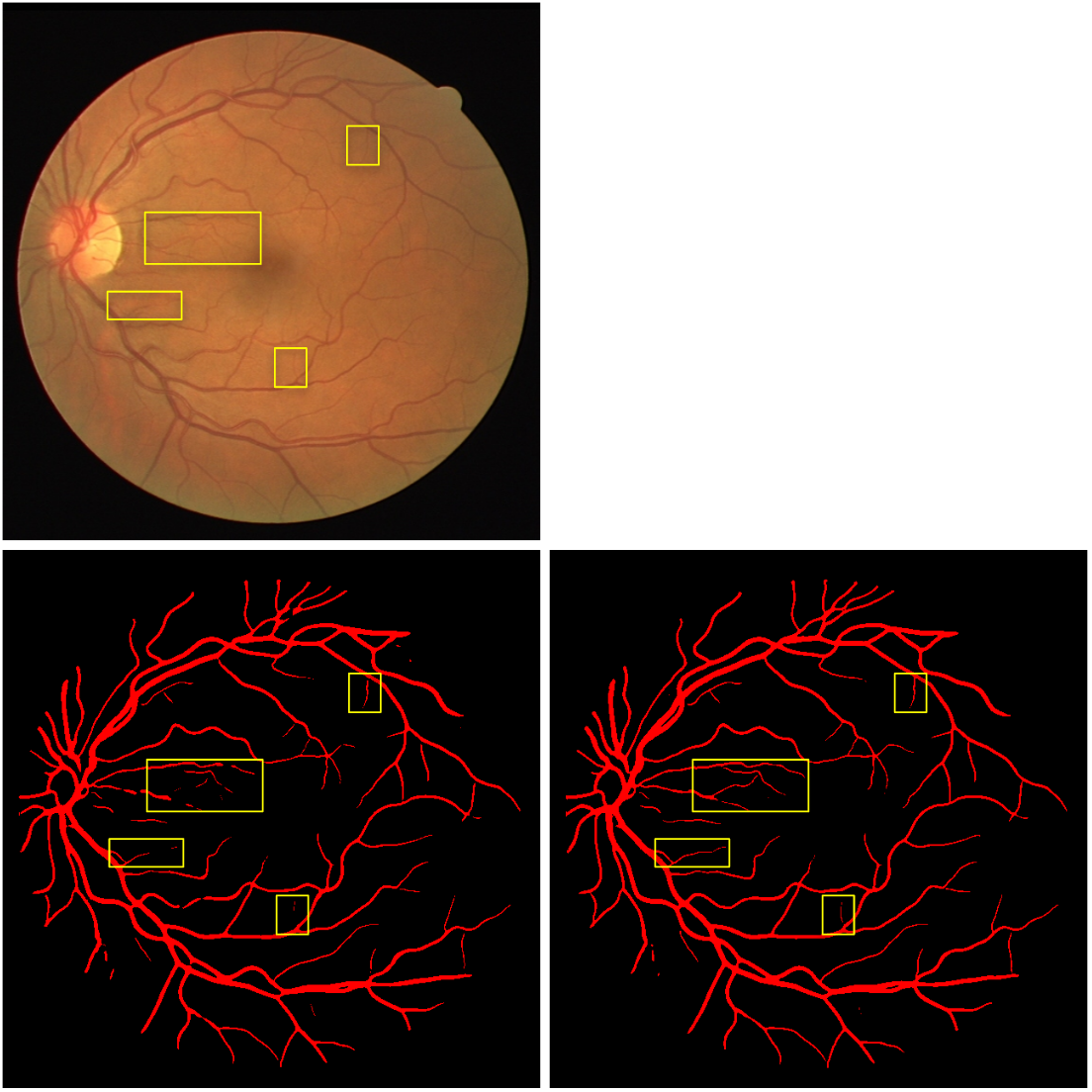}
    \caption{{\color{blue}[\emph{Chase} $\rightarrow$ \emph{Drive}]} Effectiveness of ErrorNet for cross-dataset evaluation, where the training set comes from the \emph{Chase} dataset but the test set comes from the \emph{Drive} dataset. Top: Fundus image. Bottom-Left: Segmentation result for a \emph{Drive} dataset image from the segmentation network (before error correction). Bottom-Right: corresponding segmentation result generated by ErrorNet after error correction. The yellow boxes indicate example regions where the ErrorNet model has connected fragmented vessels or sharpened vessel structures. } 
\end{figure*}

\begin{figure*}[h]
\centering
    \includegraphics[width=0.99\linewidth, keepaspectratio]{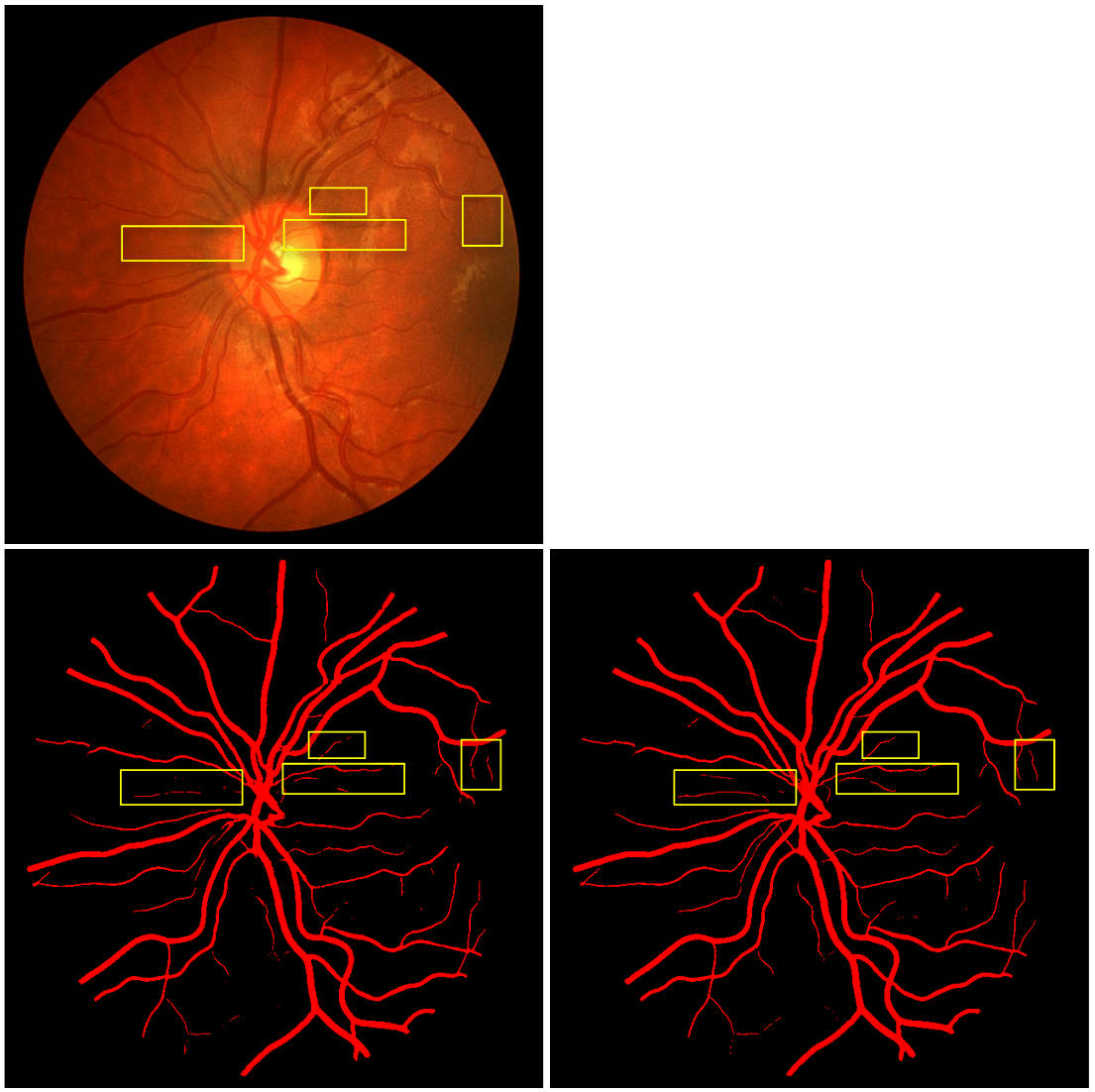}
    \caption{{\color{blue}[\emph{Chase} $\rightarrow$ \emph{Chase}]} Effectiveness of ErrorNet for same-dataset evaluation, where the training and test sets both come from the \emph{Chase} dataset. Top: Fundus image. Bottom-Left: Segmentation result for a \emph{Chase} dataset image from the segmentation network (before error correction). Bottom-Right: corresponding segmentation result generated by ErrorNet after error correction. The yellow boxes indicate example regions where the ErrorNet model has connected fragmented vessels or sharpened vessel structures. As expected, improvement is not as drastic as that of cross-dataset evaluation.} 
\end{figure*}

\begin{figure*}[h]
\centering
    \includegraphics[width=0.99\linewidth, keepaspectratio]{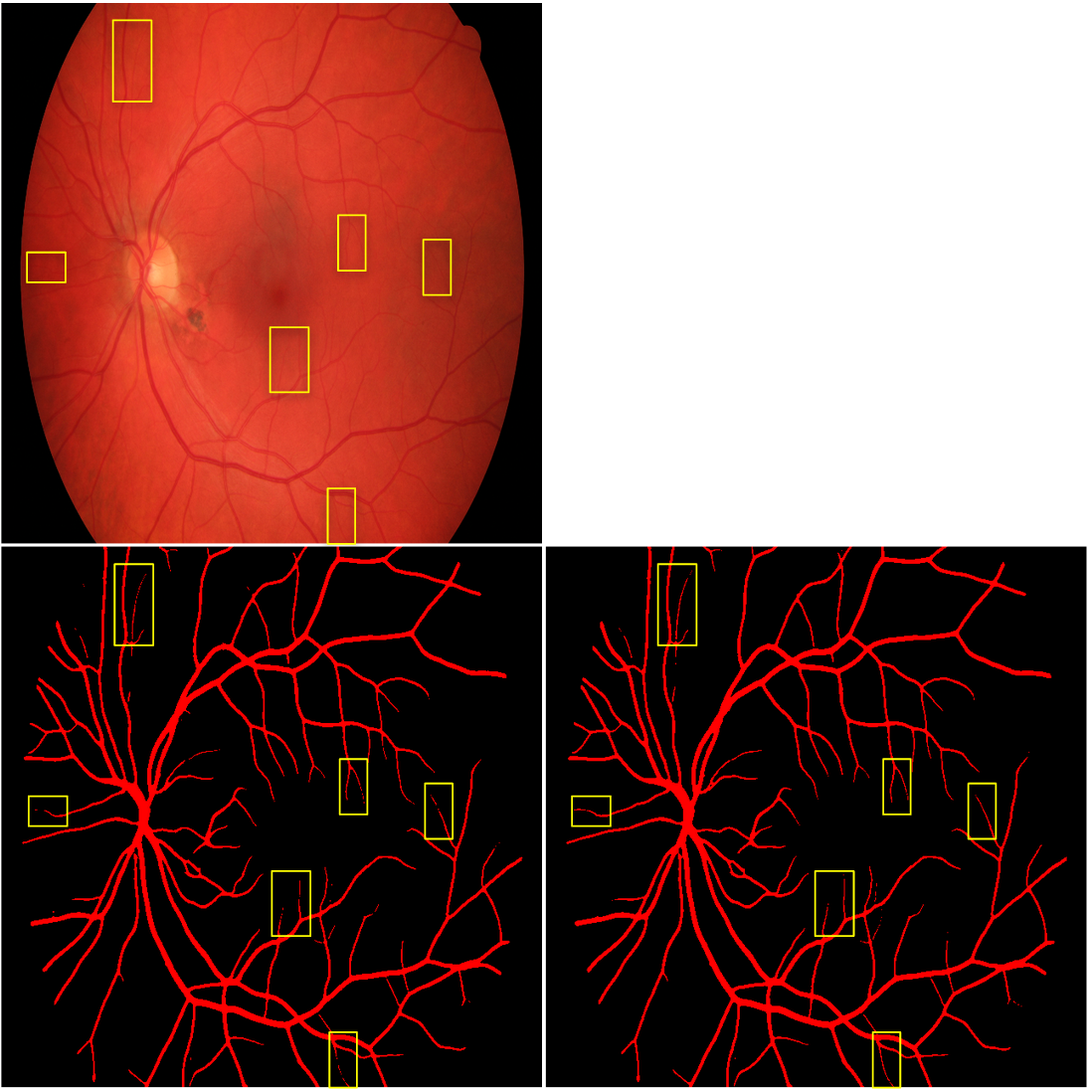}
    \caption{{\color{blue}[\emph{Aria} $\rightarrow$ \emph{HRF}]} Effectiveness of ErrorNet for cross-dataset evaluation, where the training set comes from the \emph{Aria} dataset but the test set comes from the \emph{HRF} dataset. Top: Fundus image. Bottom-Left: Segmentation result for an \emph{HRF} dataset image from the segmentation network (before error correction). Bottom-Right: corresponding segmentation result generated by ErrorNet after error correction. The yellow boxes indicate example regions where the ErrorNet model has connected fragmented vessels or sharpened vessel structures.} 
\end{figure*}

\begin{figure*}[h]
\centering
    \includegraphics[width=0.99\linewidth, keepaspectratio]{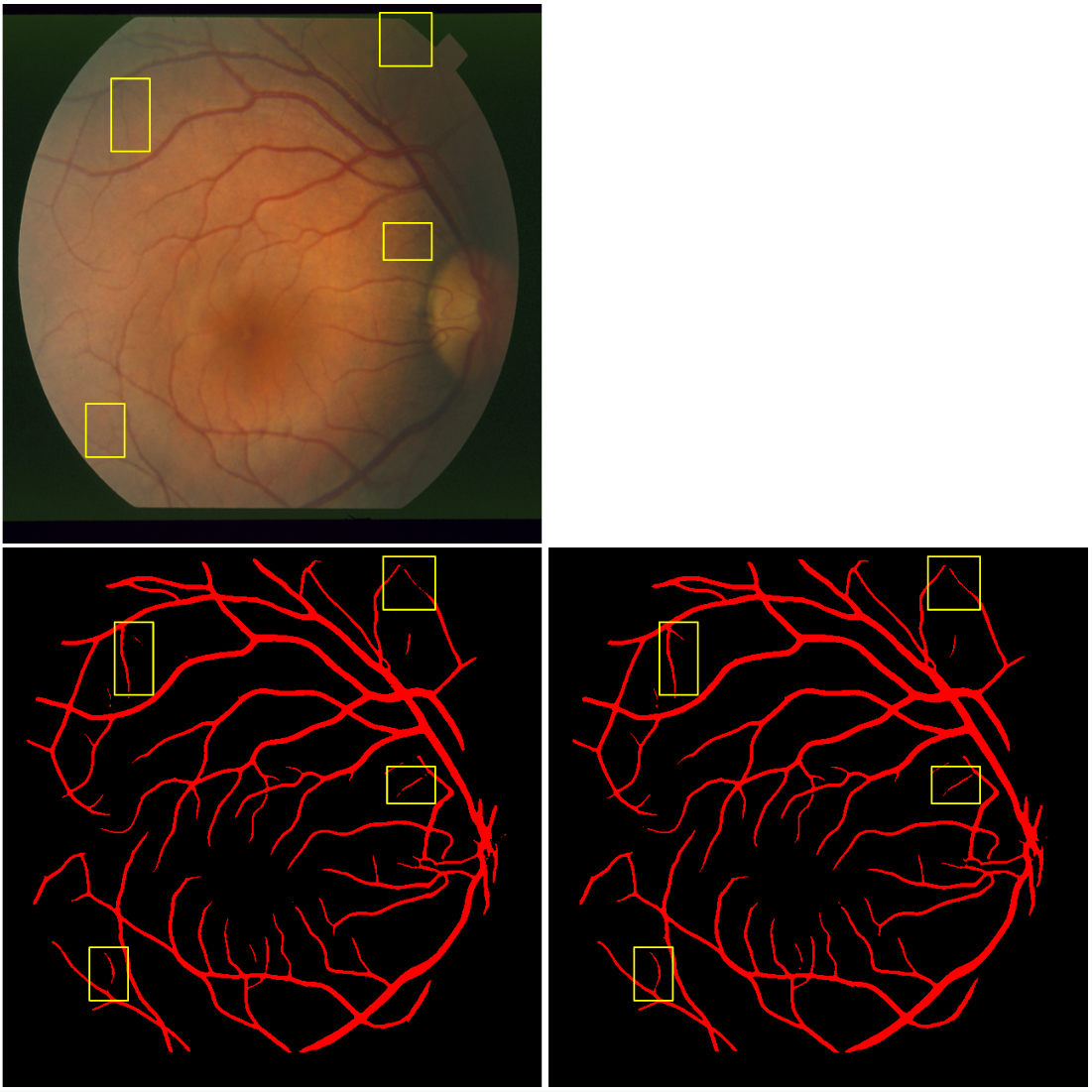}
    \caption{{\color{blue}[\emph{Aria} $\rightarrow$ \emph{Stare}]} Effectiveness of ErrorNet for cross-dataset evaluation, where the training set comes from the \emph{Aria} dataset but the test set comes from the \emph{Stare} dataset. Top: Fundus image. Bottom-Left: Segmentation result for a \emph{Stare} dataset image from the segmentation network (before error correction). Bottom-Right: corresponding segmentation result generated by ErrorNet after error correction. The yellow boxes indicate example regions where the ErrorNet model has connected fragmented vessels or sharpened vessel structures.} 
\end{figure*}

\begin{figure*}[h]
\centering
    \includegraphics[width=0.99\linewidth, keepaspectratio]{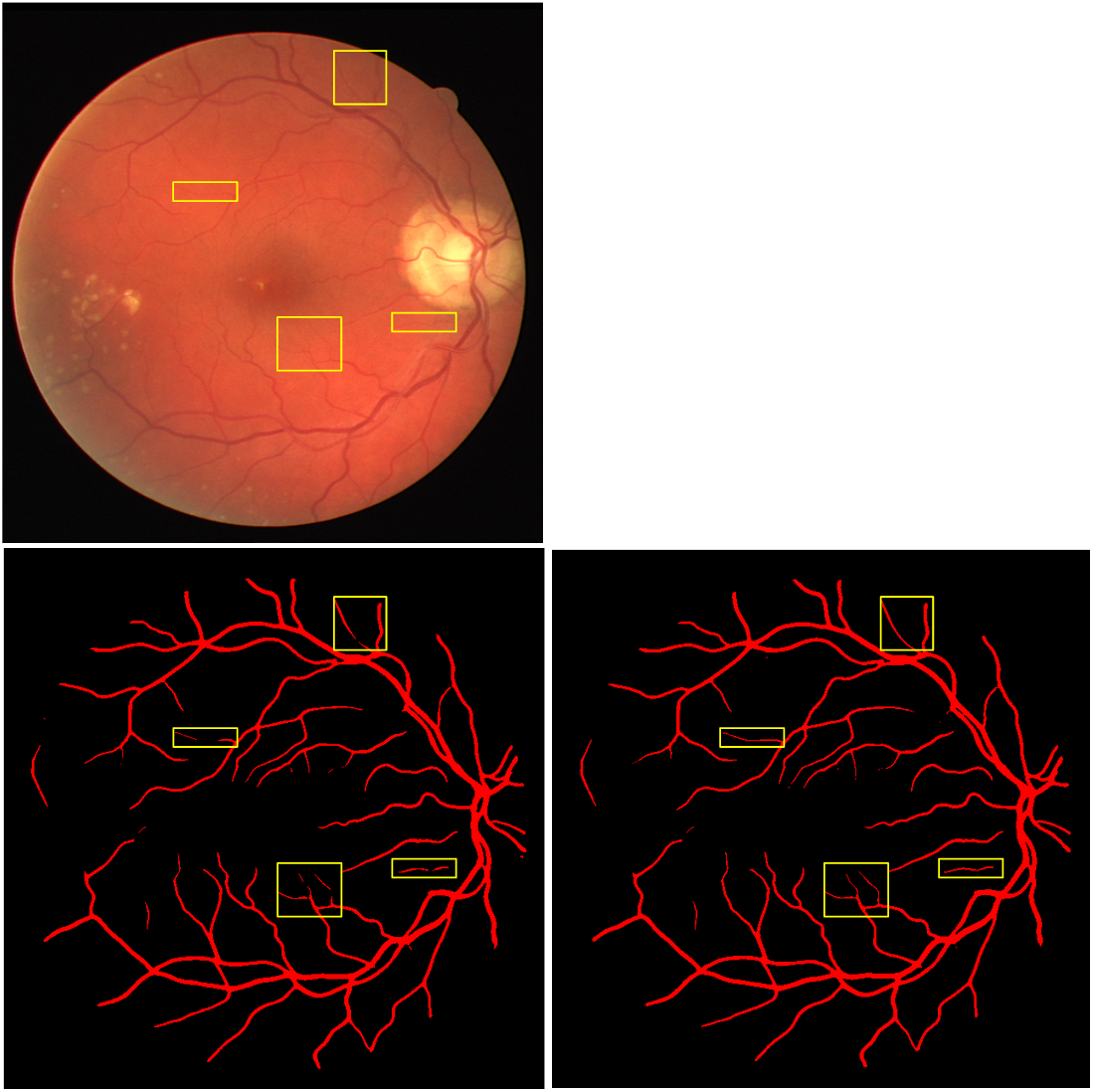}
    \caption{{\color{blue}[\emph{Aria} $\rightarrow$ \emph{Drive}]} Effectiveness of ErrorNet for cross-dataset evaluation, where the training set comes from the \emph{Aria} dataset but the test set comes from the \emph{Drive} dataset. Top: Fundus image. Bottom-Left: Segmentation result for a \emph{Drive} dataset image from the segmentation network (before error correction). Bottom-Right: corresponding segmentation result generated by ErrorNet after error correction. The yellow boxes indicate example regions where the ErrorNet model has connected fragmented vessels or sharpened vessel structures.} 
\end{figure*}

\begin{table*}[]
\caption{Architecture details for the Segmentation Network. All convolution layers use 3x3 kernels. As suggested by \cite{isensee18}, instance normalization and leaky-relu activation functions are used.}
\label{tab:SN}
\resizebox{\textwidth}{!}{%
\begin{tabular}{c|ccc}
\hline
 & Name & Feature maps (input) & Feature maps (output) \\ \hline
\multirow{14}{*}{Encoder pathway} & Conv layer - 1a & 640 x 640 x 1 & 640 x 640 x 32 \\
 & Conv layer - 1b & 640 x 640 x 32 & 640 x 640 x 32 \\
 & Max pool - 1 & 640 x 640 x 32 & 320 x 320 x 32 \\
 & Conv layer - 2a & 320 x 320 x 32 & 320 x 320 x 64 \\
 & Conv layer - 2b & 320 x 320 x 64 & 320 x 320 x 64 \\
 & Max pool - 2 & 320 x 320 x 64 & 160 x 160 x 64 \\
 & Conv layer - 3a & 160 x 160 x 64 & 160 x 160 x 128 \\
 & Conv layer - 3b & 160 x 160 x 128 & 160 x 160 x 128 \\
 & Max pool - 3 & 160 x 160 x 128 & 80 x 80 x 128 \\
 & Conv layer - 4a & 80 x 80 x 128 & 80 x 80 x 256 \\
 & Conv layer - 4b & 80 x 80 x 256 & 80 x 80 x 256 \\
 & Max pool - 4 & 80 x 80 x 256 & 40 x 40 x 256 \\
 & Conv layer -5a & 40 x 40 x 256 & 40 x 40 x 512 \\
 & Conv layer - 5b & 40 x 40 x 512 & 40 x 40 x 512 \\ \hline
\multirow{21}{*}{Decoder Pathway} & Upsample - 1 & 40 x 40 x 512 & 80 x 80 x 512 \\
 & \multirow{2}{*}{Concat - 1} & 80 x 80 x 512 (up sample - 1) & \multirow{2}{*}{80 x 80 x 768} \\
 &  & 80 x 80 x 256 (conv - 4b) &  \\
 & Conv layer - 6a & 80 x 80 x 768 & 80 x 80 x 256 \\
 & Conv layer - 6b & 80 x 80 x 256 & 80 x 80 x 256 \\
 & Upsample - 2 & 80 x 80 x 256 & 160 x 160 x 256 \\
 & \multirow{2}{*}{Concat - 2} & 160 x 160 x 256 (upsample - 2 ) & \multirow{2}{*}{160 x 160 x 384} \\
 &  & 160 x 160 x 128 (conv - 3b) &  \\
 & Conv layer - 7a & 160 x 160 x 384 & 160 x 160 x 128 \\
 & Conv layer - 7b & 160 x 160 x 128 & 160 x 160 x 128 \\
 & Upsample - 3 & 160 x 160 x 128 & 320 x 320 x 128 \\
 & \multirow{2}{*}{Concat - 3} & 320 x 320 x 128 (upsample - 3) & \multirow{2}{*}{320 x 320 x 192} \\
 &  & 320 x 320 x 64 (conv - 2b) &  \\
 & Conv layer - 8a & 320 x 320 x 192 & 320 x 320 x 64 \\
 & Conv layer - 8b & 320 x 320 x 64 & 320 x 320 x 64 \\
 & Upsample - 4 & 320 x 320 x 64 & 640 x 640 x 64 \\
 & \multirow{2}{*}{Concat - 4} & 640 x 640 x 64 (upsample - 4) & \multirow{2}{*}{640 x 640 x 96} \\
 &  & 640 x 640 x 32 (conv - 1b) &  \\
 & Conv layer - 9a & 640 x 640 x 96 & 640 x 640 x 32 \\
 & Conv layer - 9b & 640 x 640 x 32 & 640 x 640 x 32 \\
 & Output layer & 640 x 640 x 32 & 640 x 640 x 1 \\ \hline
\end{tabular}%
}
\end{table*}
\begin{table*}[]
\caption{Architecture details for the Error Injection Network. All convolution layers use 3x3 kernels. Batch normalization and relu activation functions are used throughout the network.}
\label{tab:my-table}
\resizebox{\textwidth}{!}{%
\begin{tabular}{c|ccc}
\hline
 & Name & Feature maps (input) & Feature maps (output) \\ \hline
\multirow{13}{*}{Encoder Pathway} & Conv layer - 1a & 640 x 640 x 1 & 640 x 640 x 32 \\
 & Conv layer - 1b & 640 x 640 x 32 & 640 x 640 x 32 \\
 & Max pool - 1 & 640 x 640 x 32 & 320 x 320 x 32 \\
 & Conv layer - 2a & 320 x 320 x 32 & 320 x 320 x 64 \\
 & Conv layer - 2b & 320 x 320 x 64 & 320 x 320 x 64 \\
 & Max pool - 2 & 320 x 320 x 64 & 160 x 160 x 64 \\
 & Conv layer - 3a & 160 x 160 x 64 & 160 x 160 x 128 \\
 & Conv layer - 3b & 160 x 160 x 128 & 160 x 160 x 128 \\
 & Max pool - 3 & 160 x 160 x 128 & 80 x 80 x 128 \\
 & encoder conv - 4a & 80 x 80 x 128 & 80 x 80 x 512 \\
 & encoder conv - 4b & 80 x 80 x 512 & 80 x 80 x 1 \\
 & encoder dense - mu & 80 x 80 x 1 & 6400 \\
 & encoder dense - sigma & 80 x 80 x 1 & 6400 \\ \hline
\multirow{3}{*}{VAE latent space sampling} & \multirow{2}{*}{sampling - 1} & 6400 (encoder dense - mu) & \multirow{2}{*}{6400} \\
 &  & 6400 (encoder dense - sigma) &  \\
 & reshape - 1 & 6400 & 80 x 80 x 1 \\ \hline
\multirow{11}{*}{Decoder Pathway} & Conv transpose - 1 & 80 x 80 x 1 & 160 x 160 x 64 \\
 & Conv layer - 5a & 160 x 160 x 64 & 160 x 160 x 64 \\
 & Conv layer - 5b & 160 x 160 x 64 & 160 x 160 x 64 \\
 & Conv transpose - 2 & 160 x 160 x 64 & 320 x 320 x 32 \\
 & Conv layer - 6a & 320 x 320 x 32 & 320 x 320 x 32 \\
 & Conv layer - 6b & 320 x 320 x 32 & 320 x 320 x 32 \\
 & Upsample - 3 & 320 x 320 x 32 & 640 x 640 x 32 \\
 & Conv layer - 7a & 640 x 640 x 32 & 640 x 640 x 32 \\
 & Conv layer - 7b & 640 x 640 x 32 & 640 x 640 x 32 \\
 & Output layer & 640 x 640 x 32 & 640 x 640 x 2 \\
 & Sigmoid layer & 640 x 640 x 2 & 640 x 640 x 2 \\ \hline
\end{tabular}%
}
\end{table*}

\begin{table*}[]
\caption{Architecture details for the Error Prediction Network. All convolution layers use 3x3 kernels. Batch normalization and relu activation functions are used throughout the network.}
\label{tab:EPN}
\resizebox{\textwidth}{!}{%
\begin{tabular}{c|ccc}
\hline
 & Name & Feature maps (input) & Feature maps (output) \\ \hline
\multirow{13}{*}{Encoder Pathway} & \multicolumn{1}{l}{\multirow{2}{*}{Concat - input}} & \multicolumn{1}{l}{640 x 640 x 1 (input image)} & \multicolumn{1}{l}{\multirow{2}{*}{640 x 640 x 2}} \\
 & \multicolumn{1}{l}{} & \multicolumn{1}{l}{640 x 640 x 1 (degraded segmentation)} & \multicolumn{1}{l}{} \\
 & Conv layer - 1a & 640 x 640 x 2 & 640 x 640 x 32 \\
 & Conv layer - 1b & 640 x 640 x 32 & 640 x 640 x 32 \\
 & Max pool - 1 & 640 x 640 x 32 & 320 x 320 x 32 \\
 & Conv layer - 2a & 320 x 320 x 32 & 320 x 320 x 64 \\
 & Conv layer - 2b & 320 x 320 x 64 & 320 x 320 x 64 \\
 & Max pool - 2 & 320 x 320 x 64 & 160 x 160 x 64 \\
 & Conv layer - 3a & 160 x 160 x 64 & 160 x 160 x 128 \\
 & Conv layer - 3b & 160 x 160 x 128 & 160 x 160 x 128 \\
 & Max pool - 3 & 160 x 160 x 128 & 80 x 80 x 128 \\
 & Conv layer - 4a & 80 x 80 x 128 & 80 x 80 x 256 \\
 & Conv layer - 4b & 80 x 80 x 256 & 80 x 80 x 256 \\ \hline
\multirow{16}{*}{Decoder Pathway} & Upsample - 2 & 80 x 80 x 256 & 160 x 160 x 256 \\
 & \multirow{2}{*}{Concat - 2} & 160 x 160 x 256 (upsample - 2 ) & \multirow{2}{*}{160 x 160 x 384} \\
 &  & 160 x 160 x 128 (conv - 3b) &  \\
 & Conv layer - 7a & 160 x 160 x 384 & 160 x 160 x 128 \\
 & Conv layer - 7b & 160 x 160 x 128 & 160 x 160 x 128 \\
 & Upsample - 3 & 160 x 160 x 128 & 320 x 320 x 128 \\
 & \multirow{2}{*}{Concat - 3} & 320 x 320 x 128 (upsample - 3) & \multirow{2}{*}{320 x 320 x 192} \\
 &  & 320 x 320 x 64 (conv - 2b) &  \\
 & Conv layer - 8a & 320 x 320 x 192 & 320 x 320 x 64 \\
 & Conv layer - 8b & 320 x 320 x 64 & 320 x 320 x 64 \\
 & Upsample - 4 & 320 x 320 x 64 & 640 x 640 x 64 \\
 & \multirow{2}{*}{Concat - 4} & 640 x 640 x 64 (upsample - 4) & \multirow{2}{*}{640 x 640 x 96} \\
 &  & 640 x 640 x 32 (conv - 1b) &  \\
 & Conv layer - 9a & 640 x 640 x 96 & 640 x 640 x 32 \\
 & Conv layer - 9b & 640 x 640 x 32 & 640 x 640 x 32 \\
 & Output layer & 640 x 640 x 32 & 640 x 640 x 1 \\ \hline
\end{tabular}%
}
\end{table*}

\begin{table*}[t]
\caption{\small This table is similar to Table~\ref{tab:results} in the main text {\bf with the difference being Dice is replaced with IoU for comparison}. Comparing IoU- and Dice-based results shows that the winner in each category remains unchanged (highlighted in bold). As before, grey columns indicate same-domain evaluation whereas the other columns contain the results for cross-domain evaluation. ErrorNet outperforms the competing baselines on-average, but the performance gap is wider in the presence of domain shift (cross-domain evaluation) and small sample size (the small \emph{Chase} dataset used for training). Ablation studies show that VAE and joint-training are effective in improving the performance of ErrorNet.}
\label{tab:results-dice}
\resizebox{\textwidth}{!}{%
\begin{tabular}{|l|l|l|l|lllllll|lllllll|}
\hline
\multicolumn{1}{|c|}{} & \rotatebox[origin=c]{90}{Err Pred} & \rotatebox[origin=c]{90}{VAE} & \rotatebox[origin=c]{90}{Joint Tr} & \multicolumn{1}{c}{Train on $\rightarrow$} & \multicolumn{5}{c}{CHASE} &  & \multicolumn{1}{c}{} & \multicolumn{5}{c}{ARIA} &  \\ \cline{2-18} 
\multicolumn{1}{|c|}{\multirow{-2}{*}{Architecture}} &  &  &  & \multicolumn{1}{c}{Test on $\rightarrow$} & \multicolumn{1}{c}{\cellcolor[HTML]{EFEFEF}CHASE} & \multicolumn{1}{c}{DRIVE} & \multicolumn{1}{c}{ARIA} & \multicolumn{1}{c}{STARE} & \multicolumn{1}{c|}{HRF} & Avg. & \multicolumn{1}{c}{} & \multicolumn{1}{c}{CHASE} & \multicolumn{1}{c}{DRIVE} & \multicolumn{1}{c}{\cellcolor[HTML]{EFEFEF}ARIA} & \multicolumn{1}{c}{STARE} & \multicolumn{1}{c|}{HRF} & Avg. \\ \hline
U-Net \cite{isensee18} &  &  &  &  & \cellcolor[HTML]{EFEFEF}65.7 & 51.1 & 43.2 & 42.3 & \multicolumn{1}{l|}{44.4} & 49.3 &  & 62.2 & 63.0 & \cellcolor[HTML]{EFEFEF}56.2 & 55.4 & \multicolumn{1}{l|}{56.6} & 58.6 \\ \hline
U-Net \cite{isensee18} + CRF &  &  &  &  & \cellcolor[HTML]{EFEFEF}68.4 & 48.6 & 45.6 & 39.3 & \multicolumn{1}{l|}{46.6} & 49.7 &  & {\bf 64.5} & 53.3 & \cellcolor[HTML]{EFEFEF}57.5 & 47.7 & \multicolumn{1}{l|}{{\bf 58.1}} & 56.2 \\ \hline
V-GAN\cite{son2017retinal} &  &  &  &  & \cellcolor[HTML]{EFEFEF}66.3 & 55.6 & 47.3 & 43.9 & \multicolumn{1}{l|}{49.7} & 52.5 &  & 52.3 & 61.0 & \cellcolor[HTML]{EFEFEF}53.7 & 49.5 & \multicolumn{1}{l|}{53.0} & 53.9 \\ \hline
DA-ADV \cite{dong2018unsupervised} &  &  &  &  & \cellcolor[HTML]{EFEFEF}56.6 & 53.0 & {\bf 51.7} & 47.8 & \multicolumn{1}{l|}{50.8} & 52.0 &  & 55.6 & 57.4 & \cellcolor[HTML]{EFEFEF}{\bf 57.7} & 55.4 & \multicolumn{1}{l|}{54.7} & 56.2 \\ \hline \hline
  & \ding{51} &  &  &  & \cellcolor[HTML]{EFEFEF}66.9 & 52.2 & 43.6 & 43.1 & \multicolumn{1}{l|}{45.7} & 50.3 &  & 62.3 & 56.4 & \cellcolor[HTML]{EFEFEF}56.2 & 56.1 & \multicolumn{1}{l|}{56.5} & 57.5 \\ \cline{2-18} 
 & \ding{51} & \ding{51} &  &  & \cellcolor[HTML]{EFEFEF}66.8 & 56.0 & 44.0 & 42.7 & \multicolumn{1}{l|}{50.6} & 52.0 &  & 61.6 & 63.0 & \cellcolor[HTML]{EFEFEF}56.5 & 56.5 & \multicolumn{1}{l|}{57.2} & 59.0 \\ \cline{2-18} 
\multirow{-3}{*}{ErrorNet w/ ablation} & \ding{51} & \ding{51} & \ding{51} &  & \cellcolor[HTML]{EFEFEF}{\bf 68.8} & {\bf 57.7} & 49.8 & {\bf 48.4} & \multicolumn{1}{l|}{{\bf 52.2}} & {\bf 55.3} &  & 62.2 & {\bf 65.2} & \cellcolor[HTML]{EFEFEF}56.2 & {\bf 58.7} & \multicolumn{1}{l|}{57.0} & {\bf 59.9} \\ \hline
\end{tabular}%
}
\end{table*}

\end{document}